\documentclass{aa} 
\bibpunct{(}{)}{;}{a}{}{,} 
\usepackage{graphicx}
\usepackage{txfonts}
\usepackage{amssymb,amsmath,psfig,times}
\usepackage{natbib}
\usepackage{booktabs}
\usepackage{longtable}
\def\gsim{ \lower .75ex \hbox{$\sim$} \llap{\raise .27ex \hbox{$>$}} }
\def\lsim{ \lower .75ex\hbox{$\sim$} \llap{\raise .27ex \hbox{$<$}} }

\def\beq{\begin{equation}}
\def\eeq{\end{equation}}

\def\sw{{\it Swift}}

\def\ba{BATSE}

\def\ep{$E_{\rm p}$}

\def\epo{$E^{\rm obs}_{\rm p}$}

\def\liso{$L_{\rm iso}$}

\def\flum{$\phi(L)$}
\def\flumdev{$\phi(L_0)$}

\def\gfr{$\psi(z)$}
\begin{document} 

\title{The rate and luminosity function of long Gamma Ray Bursts}


\author{
A. Pescalli 
\inst{1}\fnmsep\inst{2}\fnmsep\thanks{E--mail:alessio.pescalli@brera.inaf.it}
\and
 G. Ghirlanda \inst{2} \and R. Salvaterra \inst{3} \and G. Ghisellini \inst{2} \and
        S. D. Vergani \inst{2}\fnmsep\inst{4} \and F. Nappo \inst{1}\fnmsep\inst{2}
        \and O. S. Salafia \inst{5}\fnmsep\inst{2} \and A. Melandri \inst{1}
        \and S. Covino \inst{1} \and D. G\"otz \inst{6}
        }

   \institute{$^{1}$Universit\'a degli Studi dell'Insubria, via Valleggio 11, I-22100 Como, Italy\\
             $^{2}$INAF -- Osservatorio Astronomico di Brera, via E. Bianchi 46, I-23807 Merate, Italy\\
             $^{3}$INAF - IASF Milano, via E. Bassini 15, I-20133 Milano, Italy \\
             $^{4}$GEPI, Observatoire de Paris, CNRS, Univ. Paris Diderot, 5 place Jule Janssen, F-92190 Meudon, France\\
             $^{5}$Universit\'a degli Studi di Milano-Bicocca, Piazza della Scienza 3, I-20126 Milano, Italy\\
             $^{6}$AIM (UMR 7158 CEA/DSM-CNRS-Universit\'e Paris Diderot) Irfu/Service d'Astrophysique, Saclay, F-91191 Gif-sur-Yvette Cedex, France\\
             }


\abstract{We derive, adopting a direct method, the luminosity function and the formation rate of long Gamma Ray Bursts through a complete, flux--limited, sample of \sw\ bursts which has a high level of completeness in redshift $z$ ($\sim 82\% $). We parametrise the redshift evolution of the GRB luminosity as $L=L_{0} (1+z)^k$ and we derive $k = 2.5$,  consistently with recent estimates. The de-evolved luminosity function \flumdev\ of GRBs can be represented by a broken power law with slopes $a=-1.32 \pm 0.21$ and $b=-1.84 \pm 0.24$ below and above, respectively, a characteristic break luminosity $L_{\rm 0,b}=10^{51.45 \pm 0.15}$ erg/s. Under the hypothesis of luminosity evolution we find that the GRB formation rate increases with redshift up to $z\sim 2$, where it peaks, and then decreases in agreement with the shape of the cosmic star formation rate. We test the direct method through numerical simulations and we show that if it is applied to incomplete (both in redshift and/or flux) GRB samples it can misleadingly result in an excess of the GRB formation rate at low redshifts.   } 

          
\maketitle

\section{Introduction}

Since the discovery of Gamma Ray Bursts (GRBs), one of the most important questions was related to their distance scale (i.e. whether galactic or cosmological) which had immediate implications on their associated luminosities and energetics. Through the afterglow detection (Costa et al. 1997; Van Paradijs 1997) and first redshift measurements, GRBs were proven to be cosmological sources with large isotropic equivalent luminosities exceeding, in few cases, $10^{54}$ erg s$^{-1}$.  The  pinpointing of the GRB afterglow, made available by the fast slewing of the \sw\ satellite (Gehrels et al. 2004), coupled with intense efforts to acquire early time optical spectra from ground, allowed us to measure the redshifts $z$ of GRBs with an average efficiency of 30\%. 
Among these, GRB\,090423 (with a spectroscopic $z=8.2$ - Salvaterra et al. 2009a; Tanvir et al. 2009) and GRB\,090429B (with photometric redshift $z=9.4$ - Cucchiara et al. 2011) represent the farthest objects of stellar origin known so far. 

Two of the key properties characterising the population of GRBs are (a) their cosmic rate \gfr\ (GRB formation rate, GRBFR hereafter), representing the number of bursts per unit comoving volume and time as a function of redshift, and (b) their luminosity function \flum\ (LF hereafter), representing the relative fraction of bursts with a certain luminosity. Here, with $\phi(L)$ we refer to the differential luminosity function defined as $dN(L)/dL$. 

Recovering \gfr\ and \flum\ of GRBs allows us to test the nature of their progenitor (e.g. through the comparison with the  cosmic star formation rate) and to study the possible presence of sub--classes of GRBs at the low end of the luminosity function (e.g. Liang et al. 2007, see also Pescalli et al. 2015). 
These two functions have been derived for the population of long GRBs (e.g. Daigne et al. 2006; Guetta \& Della Valle 2007; Firmani et al. 2004; Salvaterra \& Chincarini 2007; Salvaterra et al. 2009b, 2012; Wanderman \& Piran 2010; Yu et al. 2015; Petrosian et al. 2015) through different methods and samples of bursts (\S2). For the population of short GRBs, \gfr\ and \flum\ have been less securely constrained (e.g. Guetta \& Piran 2005, 2006; Nakar 2006; Berger et al. 2014; D'Avanzo et al. 2015) due to the limited number of bursts with measured redshifts. 

However, \gfr\ and \flum\ cannot be derived straightforwardly using all GRBs with known redshift since these samples are affected by  observational biases. Specific methods that correct for  such biases should be adopted. The main approaches that  have been used so far (\S2) agree on the shape of the luminosity function (typically represented by a broken power law) but lead to remarkably different results on the cosmic GRB rate (particularly at low redshifts). Independently from the method used to recover these two functions, most of the previous studies (see however Salvaterra et al. 2012) adopted either heterogeneous samples (i.e. including GRBs detected by different satellites/GRB detectors which have different sensitivities) and/or incomplete samples. In particular, incompleteness is induced by several effects such as the variation of the trigger efficiency and the redshift measurement. Accounting for such instrumental effects is extremely difficult in practice. 

An alternative is to work with complete samples, at the expense of the number of GRBs in the sample. Salvaterra et al. (2012) (S12) defined a complete flux--limited sample of GRBs (called BAT6) detected by \sw\ which, despite containing a relatively small number of GRBs, has a high redshift completeness and has been extensively used to test various prompt and afterglow properties of GRBs in an unbiased way (Campana et al. 2012, D'Avanzo et al. 2012, Ghirlanda et al. 2012, Melandri et al. 2012, Nava et al. 2012, Covino et al. 2013, Melandri et al. 2014, Vergani et al. 2014). 

The aim of this work is to derive \flum\ and \gfr\ through this complete sample of GRBs detected by \sw. We summarise the main different approaches that have been used in the literature to derive the luminosity function and the formation rate of GRBs (\S2) and present the updated sample used in this work in \S3. We adopt the $C^{-}$ direct method (Lynden-Bell et al. 1971) to derive the \flum\ and \gfr\ and compare it with previous results in \S4. Throughout the paper we assume a standard $\Lambda$CDM cosmological model with $\Omega_{m}=0.3$ and $\Omega_{\Lambda}=0.7$ with $H_{0}=70$ km s$^{-1}$ Mpc$^{-1}$. We use the symbol $L$ to indicate the isotropic equivalent luminosity \liso\, omitting for simplicity the subscript ``iso".

\section{{\bf \flum} and {\bf \gfr} of long GRBs}

The number of GRBs detectable by a given instrument above its sensitivity flux limit $S$ can be expressed as: 

\begin{equation}
N(>S) = \frac{\Omega T}{4\pi} \int ^{z(L_{\rm max},S)} _{0} \int ^{L_{\rm max}}_{L_{\rm lim}(S,z)} \phi(L,z)   \frac{\psi(z)}{(1+z)} \frac{dV}{dz} \,dL\,dz 
\label{eq1}
\end{equation}

where $\Omega$ and $T$ are the instrument field of view and time of operation, respectively, and $dV/dz$ is the differential comoving volume. Here, $z(L_{\rm max},S)$ is the maximum redshift at which a burst with  $L_{\rm max}$ would still be above the instrumental flux limit $S$; 
$L_{\rm lim}(S,z)$ is the minimum observable luminosity as a function of z (i.e. that corresponds to a flux above $S$). 

If \flum\ and \gfr\ are known, it is possible to derive from Eq. \ref{eq1} the flux distribution of the population of GRBs observable by a given detector, knowing its instrumental parameters. By reversing this argument, one can assume the functional forms of \flum\ and \gfr\ (e.g. specified through a set of free parameters) and constrain them by fitting the the model flux distribution (i.e. $N(>S)$) to the observed flux distribution of a given instrument. This \emph{indirect} method has been used to infer the luminosity function (e.g. Firmani et al. 2004; Salvaterra \& Chincarini 2007; Salvaterra et al. 2009, 2012) by fitting e.g. the flux distribution of the large population of GRBs detected by \ba. 

The number of free parameters, if both \flum\ and \gfr\ are to be constrained, can be large. One possibility is to assume that, based on the massive star progenitor origin of long bursts, the GRB cosmic rate traces the cosmic star formation rate, i.e. \gfr\ $\propto\psi_{\star}(z)$. In this way, the method allows to derive the free parameters of \flum\ by fitting the result of Eq. \ref{eq1} to large, statistically significant, samples of observed GRBs. The assumed \gfr\ can be tested by fitting the observed redshift distribution of a sample of bursts with measured $z$. 

In the simplest scenario, the two functions \flum\ and \gfr\ are independent. However, more realistic analyses also considered the possible evolution of either the luminosity function or the GRB formation rate with redshift.  For example, in the case of luminosity evolution, the burst luminosity depends on $z$ through the relation $L(z) = L_{0}(1+z)^{\rm k}$ (luminosity evolution scenario). Alternatively, the GRB formation rate \gfr\ varies with redshift with a similar analytical dependence $\psi(z) \propto \psi_{\star}(z)(1+z)^{\rm d}$ (density evolution scenario). This means that the progenitor characteristics evolve with $z$ and that the ratio of the GRB formation rate to the cosmic star formation rate is not constant.  Among the drawbacks of this method is that it relies on the assumption of a specific functional form of \flum\ (and simple functions, e.g. power law, broken power law or power law with a cutoff at low luminosities, have been adopted) and it  allows to test only for evolution of the luminosity or of the rate independently.

Salvaterra et al. (2012) applied the indirect method to a complete sample of GRBs detected by \sw\ (\S3). They find that either a luminosity evolution with $k = 2.1 \pm 0.6$ or a density evolution with $d = 1.7 \pm 0.5$ can reproduce the flux distribution of \ba\ bursts and the redshift distribution of the \sw\ complete sample. However, they can not discriminate between these two scenarios. They derive the luminosity function \flum\ testing two analytical models: a power-law with an exponential cut-off at low luminosities and a broken power-law (BPL, $\phi(L) \propto (L/L_{\rm b})^{a,b}$, where $a$ and $b$ are the slopes of the power law below and above the break $L_{\rm b}$) adopting a minimum GRB luminosity $L_{\rm min}=10^{49}$ erg/s. For the BPL model they found $a = -0.74 _{-1.42} ^{+1.36}$, $b = -1.92 _{-0.14} ^{+0.11}$, $L_{\rm b} = 5.5 ^{+6.9} _{-3.4} \times 10^{50}$ erg/s and $a = -1.5 _{-0.16} ^{+0.32}$, $b = 2.32 ^{+0.77} _{-0.32}$, $L_{\rm b} = 3.8 ^{+6.3} _{-2.7}\times 10^{52}$ erg/s in the case of  luminosity and  density evolution scenario, respectively.

The alternative method is based on the direct derivation of the \flum\ and \gfr\ from observed samples of GRBs with measured $z$ and $L$. This method has been inherited from the studies of the luminosity function of quasars and blazars (e.g. Chiang et al. 1998, Maloney \& Petrosian 1999, Singal et al. 2012,2013) and it has been applied to GRBs (Lloyd et al. 1999, Kocevski \& liang 2006). Wanderman \& Piran (2010) adopt a maximum likelihood estimator to derive the discrete luminosity function and cosmic formation rate. They use the sample of $\sim$100 GRBs detected by \sw\ with measured redshift (through optical afterglow absorption lines and photometry). Despite this sample might suffer from incompleteness, they derive \flum\ (extending from $10^{50}$ erg/s up to $10^{54}$ erg/s) which can be represented by a broken power law  with $a=-1.2$, $b=-2.4$ and $L_{\rm b}=10^{52.5}$ erg/s. Similarly, they also derived the discrete GRB formation rate \gfr\ which can be represented by a broken power-law as a function of $(1+z)$ with indices $n_{\rm 1}=2.1^{+0.5}_{-0.6}$ and $n_{\rm2}=-1.4^{+2.4}_{-1.0}$ peaking at $z = 3.1^{+0.6}_{-0.8}$. This rate is consistent with the SFR of Bouwenset al. (2009) for $z \lesssim 3$.
They assume that the luminosity is independent from redshift. 

More recently, Yu et al. (2015 - Y15 hereafter) and Petrosian et al. (2015 - P15 hereafter) applied a statistical method to reconstruct the discrete \flum\ and \gfr\ from a sample of \sw\ bursts with measured redshifts. They both find a strong luminosity evolution with $k\sim2.3$. Their results converge towards a cumulative luminosity function described by a broken power-law with $\alpha =-0.14 \pm 0.02$, $\beta =-0.7\pm0.03$, $L_{\rm b}=1.43 \times 10^{51}$ erg/s (Y15) and $\alpha=-0.5$, $\beta=-2.2$, $L_{\rm b}=10^{51}$ erg/s (P15). These indices ($\alpha$ and $\beta$) are the slopes of the cumulative luminosity function which is linked  to the differential one through the integral $\Phi(L) = N(>L) = \int _{L}^{L_{\rm max}} \phi(L) \, dL$. Therefore, for a BPL luminosity function, the slopes of the differential form are $(a,b) = (\alpha-1,\beta-1)$.

Intriguingly, they find, at odd with respect to previous works, that the GRB rate is flat or decreases from the local Universe up to $z=1$. If compared to the SFR, this behaviour would imply a relative excess of the GRB formation rate with respect to $\psi_{*}(z)$ at $z\le$1 (if both are normalised to their respective peaks). They dub this behaviour \emph{the excess of GRBs at low redshifts}. This result is puzzling also because it is completely at odds with the findings of the works based on the properties of GRB host galaxies. In fact, Vergani et al. (2014), Perley et al. (2015a, 2015b, 2015c) and Kruhler et al. (2015), performed multi-wavelength and spectroscopic studies on the properties (stellar masses, luminosities, SFR and metallicity) of GRB host galaxies of different complete GRB samples and compared them to those of the star-forming galaxies selected by galaxy surveys. All their results clearly indicate that at $z<1$ only a small fraction of the star-formation produces GRBs. 

Both P15 and Y15 apply a statistical method (Efron \& Petrosian 1992) to remove the redshift dependence of the luminosity induced by the flux-cut in the selected GRB sample. They use GRBs detected by \sw\ with measured redshifts. However, while Yu et al. (2015) work with the bolometric luminosity of GRBs, Petrosian et al. (2015) adopt the luminosity in the \sw/BAT (15-150 keV) energy band. Y15 use all GRBs detected by \sw\ with a measured redshift and well constrained spectral parameters: despite their relatively large number of objects ($\sim130$), this is an incomplete sample. P15 account for incompleteness by cutting their sample to a relatively large flux level, at the expense of the total number of bursts, i.e working with $\sim 200/250$ events with measured $z$. 

Independently from the method adopted to recover \flum\ and \gfr, one key point is the definition of the sample. S12 showed the importance of working with complete samples of GRBs (see also Hjorth et al. 2012).  Here we start with the so called BAT6 \sw\ sample (S12) and extend it with additional bursts that satisfy its selection criteria (\S3). We will then use it to derive the luminosity function and the cosmic GRB formation rate (\S4).    

\section{BAT6 extended version}
\label{sec3}

The BAT6 complete sample as defined in S12 was composed by 58 \sw\ GRBs with (i) favourable observing conditions  for their redshift measurement as proposed in Jackobsson et al. (2006) and (ii)  a peak photon flux $P \ge 2.6$ ph cm$^{-2}$ s$^{-1}$ (integrated over the 15--150 keV \sw/BAT energy band). This sample, which is complete in flux by definition,  turned out, after selection, \emph{also}  to be highly complete ($\sim$90\%) in redshift (i.e. 52/58 bursts have $z$). 

The study of the isotropic equivalent luminosity $L$ of the bursts of the BAT6 sample requires the knowledge of their broad band prompt emission spectrum. Nava et al. (2012) collected the 46/52 GRBs, within the BAT6, with measured \ep\ and $z$. Six bursts with measured $z$ did not have \ep\ measurements. One of the main drawbacks of the narrow/soft energy range of the BAT instrument is the difficulty to measure the peak \ep\ of the $\nu F_{\nu}$ spectrum for several bursts it detects. Other instruments (e.g. \textit{Konus/Wind} - Aptekar et al. 1995, \textit{Fermi}/GBM - Meegan et al. 2009 or \textit{Suzaku}/WAM - Yamaoka et al. 2009) compensate for this lack, thanks to their wide energy range, measuring a spectrum extending from few keV to several MeV. 

Sakamoto et al. (2009) showed that for \sw\ bursts with measured \ep\ there is a correlation between the slope of the spectrum $\alpha_{\rm PL}$ (when fitted with a single powerlaw model) and the peak energy \ep\ (measured by fitting a curved model). With the aim of enlarging the sample of Nava et al. (2012), we estimated \ep\ of six bursts of the BAT6, whose BAT spectrum is fitted by a single power law, through this relation (Sakamoto et al. 2009) and verified that the values obtained are consistent with those of the other bursts (we performed the K$-$S test finding a probability of $\sim 70\%$ that the two sets of peak energies originate from the same distribution). We find that all but one GRB (i.e. 070306) have $E_{\rm p}^{\rm rest}$ = \epo$(1+z)$ consistent with the upper/lower limit reported in Nava et al. (2012). Therefore, we firstly extended the BAT6 sample of Nava et al. (2012) with measured $z$ and $L$ to 50/58 bursts. 


Since the construction of the BAT6, other bursts satisfying its selection criteria were detected by \sw. Moreover, some burst already present in the original BAT6 sample were re--analysed and either their redshifts and/or their spectral properties were revised.  So our first aim was to revise the BAT6 sample. In particular, the revision of 8 redshifts is here included (marked in italics in the table - their luminosity has been updated). The revised BAT6 sample then contains 56/58 GRBs with measured $z$ and 54/58 with also a bolometric isotropic luminosity $L$. Considering only the redshift, the sample is $\sim 97\%$ complete, while if we also require the knowledge of $L$, the completeness level is only slightly smaller ($\sim93\%$).      
              
\begin{table*}
\scriptsize
\caption{{\label{tab1}} BAT6ext (BAT6 extended) GRB complete sample. Columns report, in order, the redshift $z$, the spectral photon indices $\alpha$ and $\beta$, the peak flux in units of $10^{-7}$ erg cm$^{-2}$ s$^{-1}$ (except for those with the $^\star$ which are in units of photons cm$^{-2}$ s$^{-1}$), the respective energy band, the rest-frame peak energy $E_{\rm p}$ and the bolometric equivalent isotropic luminosity $L$ (calculated in the $[1-10^{4}]$ rest-frame energy range). For $L$ we also give the $1\sigma$ error. $^{\rm b}$ Bursts with missing $z$ are reported with their observer frame $E_{\rm p}^{\rm obs}$ for completeness, they are not used in the present work. $^{\rm c}$ The peak energy has been estimated with the relation of Sakamoto et al. (2009). The eight GRBs (already present in the compilation of Nava et al. 2012) with an updated redshift estimate are marked in italics. In the last column we report the references, in order, for the spectral parameters and for the redshift: 1) Nava et al. (2012), 2) Covino et al. (2013), 3) Kruhler et al. (2015), 4) GCN\,\#\,12133, 5) GCN\,\#\,12135-12137, 6) GCN\,\#\,12190, 7) GCN\,\#\,12352, 8) GCN\,\#\,12424, 9) GCN\,\#\,12431, 10) GCN\,\#\,12749, 11) GCN\,\#\,12761, 12) GCN\,\#\,12801, 13) GCN\,\#\,12839, 14) GCN\,\#\,12874, 15) GCN\,\#\,12865, 16) GCN\,\#\,13120, 17) GCN\,\#\,13118, 18) GCN\,\#\,13412, 19) GCN\,\#\,13536, 20) GCN\,\#\,13532, 21) GCN\,\#\,13559, 22) GCN\,\#\,13562, 23) GCN\,\#\,13634, 24) GCN\,\#\,13628, 25) GCN\,\#\,13721, 26) GCN\,\#\,13723, 27) GCN\,\#\,13990, 28) GCN\,\#\,13992, 29) GCN\,\#\,13997, 30) GCN\,\#\,14052, 31) GCN\,\#\,14419, 32) GCN\,\#\,14437, 33) GCN\,\#\,14487, 34) GCN\,\#\,14491, 35) GCN\,\#\,14469, 36) GCN\,\#\,14493, 37) GCN\,\#\,14545, 38) GCN\,\#\,14575, 39) GCN\,\#\,14567, 40) GCN\,\#\,14720, 41) GCN\,\#\,14808, 42) GCN\,\#\,14796, 43) GCN\,\#\,14869, 44) GCN\,\#\,14959, 45) GCN\,\#\,14956, 46) GCN\,\#\,15064, 47) GCN\,\#\,15145, 48) GCN\,\#\,15144, 49) GCN\,\#\,15203, 50) GCN\,\#\,15187, 51) GCN\,\#\,15413, 52) GCN\,\#\,15407, 53) GCN\,\#\,15452, 54) GCN\,\#\,15450, 55) GCN\,\#\,15669, 56) GCN\,\#\,15805, 57) GCN\,\#\,15800, 58) GCN\,\#\,15853, 59) GCN\,\#\,16134, 60) GCN\,\#\,16125, 61) GCN\,\#\,16220, 62) GCN\,\#\,16217, 63) GCN\,\#\,16262, 64) GCN\,\#\,16310, 65) GCN\,\#\,16423, 66) GCN\,\#\,16473, 67) GCN\,\#\,16495, 68) GCN\,\#\,16489, 69) GCN\,\#\,16512, 70) GCN\,\#\,16505.}
\centering
\begin{tabular}{lccccccc}
\toprule
\multicolumn{1}{l}{GRB} &
\multicolumn{1}{c}{$z$} & 
\multicolumn{1}{c}{$\alpha$[$\beta$]} &
\multicolumn{1}{c}{Peak flux} &
\multicolumn{1}{c}{Range} &
\multicolumn{1}{c}{$E_{\rm p}$} &
\multicolumn{1}{c}{$L$} &
\multicolumn{1}{c}{Ref} \\
    &    &                     & $10^{-7}$ erg cm$^{-2}$ s$^{-1}$ &(keV) &(keV) &($\times 10^{ 51}$ erg/s) & \\
    &    &                     &$^\star$ phot cm$^{-2}$ s$^{-1}$ &  &  &  & \\
\midrule

050318 & $1.44$ & $-1.34\pm 0.32$ & $2.20 \pm 0.17$ & $[15-150]$ & $115 \pm 27$ & $4.76 \pm 0.86$ & 1,1 \\

050401 & 2.9 & $-1.0[-2.45]$ & $24.5 \pm 1.2$ & $[20-2000]$ & $499 \pm 117$ & $201 \pm 11$ & 1,1 \\

050416A & 0.653 & $-1.0[-3.4]$ & $5.0 \pm 0.5^{\star}$ & $[15-150]$ & $26 \pm 4$ & $0.97 \pm 0.12$ & 1,1 \\

050525A  & 0.606 & $-0.99 \pm 0.11$ & $47.7 \pm 1.2^{\star}$ & $[15-350]$ & $127 \pm 6$ & $7.24 \pm 0.28$ & 1,1 \\	 

050802$^{\rm c}$ & 1.71 & $-1.6 \pm 0.1$ & $2.21 \pm 0.35$ & $[15-150]$ & 301 & $9.51 \pm 1.71$ & 1,1 \\

050922C & 2.198 & $-0.83 \pm 0.24$ & $45 \pm 7$ & $[20-2000]$ & $416 \pm 118$ & $187 \pm 30$ & 1,1 \\

060206 &	 4.048 & $-1.12 \pm 0.30$ & $2.02 \pm 0.13$ & $[15-150]$ & $409 \pm 116$ & $49.6 \pm 7.1$ & 1,1 \\    

060210 &	 3.91 &	$-1.12 \pm 0.26$ & $2.8 \pm 0.3^{\star}$ & $[15-150]$ & $574 \pm 187$ & $52.8 \pm 11.1$ & 1,1 \\

\textit{060306} & 1.55 & $-1.2 \pm 0.5$ & $4.71 \pm 0.28$ & $[15-150]$ & $178.5 \pm 76.5$ & $11.49 \pm 2.26$ & 1,2 \\

060614 &	 0.125 & $-1.5$ & $11.6 \pm 0.7^{\star}$ & $[15-150]$ & $55 \pm 45$ & $0.05 \pm 0.01$ & 1,1 \\

060814 &	 1.92 &	$-1.43 \pm 0.16$ & $21.3 \pm 3.5$ & $[20-1000]$ & $750 \pm 245$ & $71.7 \pm 13.1$ & 1,1 \\      

060904A & $-$ & $-1.22 \pm 0.05$ & $13 \pm 3$ & $[20-10\,000]$ & $235 \pm 25^{\rm b}$ & $-$ & 1,1 \\ 	

060908 &	 1.88 & $-0.93 \pm 0.25$ & $2.81 \pm 0.23$ & $[15-150]$ & $426 \pm 207$ & $12.7 \pm 3.1$ & 1,1 \\

060912A$^{\rm c}$ & 0.94 & $-1.85 \pm 0.08$ & $25 \pm 9$ & $[20-10\,000]$ & 127 & $20.6 \pm 7.4$ & 1,1 \\

060927 & 5.47 & $-0.81 \pm 0.36$ & $2.47 \pm 0.17$ & $[15-150]$ & $459 \pm 90$ & $108.7 \pm 13.1$ & 1,1 \\

061007 &	 1.261 & $-0.75 \pm 0.02\,[-2.79 \pm 0.09]$ & $120 \pm 10$ & $[20-10\,000]$ & $965 \pm 27$ & $109.2 \pm 8.9$ & 1,1 \\

061021 &	 0.346 & $-1.22 \pm 0.13$ & $37.2 \pm 9.3$ & $[20-2000]$ & $1046 \pm 485$ & $1.77 \pm 0.46$ & 1,1 \\   

061121 &	 1.314 & $-1.32 \pm 0.05$ & $128 \pm 17$ & $[20-5000]$ & $1402 \pm 185$ & $142 \pm 19$ & 1,1 \\

061222A & 2.09 &	 $-1.00 \pm 0.05\,[-2.32 \pm 0.38]$ & $48 \pm 13$ & $[20-10\,000]$ & $1091 \pm 167$ & $140 \pm 38$ & 1,1 \\  

070306$^{\rm c}$ & 1.50 & $-1.67 \pm 0.1$ & $3.04 \pm 0.16$ & $[15-150]$ & $>263$ & $>9.99$ & 1,1 \\  

\textit{070328} & 2.063 & $-1.11 \pm 0.04\,[-2.33 \pm 0.24]$ & $59 \pm 12$ & $[20-10\,000]$ & 2349 & $157.6 \pm 37.6$ & 1,3 \\

\textit{070521} & 2.087 & $-0.93 \pm 0.12$ & $41.2 \pm 9.1$ & $[20-1000]$ & $685.6 \pm 73.6$ & $144.1 \pm 32.6$ & 1,3 \\    

071020 & 2.145 & $-0.65 \pm 0.29$ & $60.4 \pm 20.8$ & $[20-2000]$ & $1013 \pm 204$ & $213 \pm 73$ & 1,1 \\

071112C$^{\rm c}$ & 0.82 & $-1.09 \pm 0.07$ & $8.0 \pm 1.0^{\star}$ & $[15-150]$ & 596 & $6.57 \pm 0.86$ & 1,1\\

071117 & 1.331 & $-1.53 \pm 0.15$ & $66.6 \pm 18.3$ & $[20-1000]$ & $648 \pm 317$ & $95.4 \pm 28.4$ & 1,1 \\ 

080319B & 0.937 & $-0.86 \pm 0.01\,[-3.59 \pm 0.45]$ & $226 \pm 21$ & $[20-7000]$ & $1307 \pm 43$ & $101.6 \pm 9.4$ & 1,1\\      

080319C & 1.95 & $-1.20 \pm 0.10$ & $33.5 \pm 7.4$ & $[20-4000]$ & $1752 \pm 504$ & $96.1 \pm 21.7$ & 1,1\\

080413B & 1.10 & $-1.23 \pm 0.25$ & $14.0 \pm 0.6$ & $[15-150]$ & $163 \pm 34$ & $14.9 \pm 1.8$ & 1,1 \\

080430$^{\rm c}$ & 0.77 & $-1.73 \pm 0.08$ & $1.82 \pm 0.13$ & $[15-150]$ & 149 & $1.16 \pm 0.13$ & 1,1 \\

\textit{080602}$^{\rm c}$ & 1.820 & $-0.96 \pm 0.63$ & $19.2 \pm 5.8$ & $[20-1000]$ & $1216$ & $51 \pm 17$ & 1,3\\ 

080603B & 2.69 & $-1.23 \pm 0.64$ & $15.1 \pm 3.9$ & $[20-1000]$ & $376 \pm 214$ & $116.6 \pm 38.9$ & 1,1\\

080605 & 1.64 & $-1.03 \pm 0.07$ & $160 \pm 33$ & $[20-2000]$ & $665 \pm 48$ & $308.7 \pm 62.8$ & 1,1\\  

080607 & 3.036 & $-1.08 \pm 0.06$ & $269 \pm 54$ & $[20-4000]$ & $1691 \pm 169$ & $2260 \pm 446$ & 1,1\\

080613B & $-$ & $-1.05 \pm 0.18$ & $47.6 \pm 13.1$ & $[20-3000]$ & $33 \pm 239^{\rm b}$ & $-$ & 1,1\\  	

080721 & 2.591 & $-0.96 \pm 0.07\,[-2.42 \pm 0.29]$ & $211 \pm 35$ & $[20-7000]$ & $1785 \pm 223$ & $1039 \pm 173$ & 1,1\\
                             
080804 & 2.20 & $-0.72 \pm 0.04$ & $7.30 \pm 0.88$ & $[8-35\,000]$ & $810 \pm 45$ & $27.0 \pm 3.2$ & 1,1 \\

080916A & 0.689 & $-0.99 \pm 0.05$ & $4.87 \pm 0.27$ & $[8-35\,000]$ & $208 \pm 11$ & $1.08 \pm 0.06$ & 1,1\\

081007 & 0.53 & $-1.4 \pm 0.4$ & $2.2 \pm 0.2^{\star}$ & $[25-900]$ & $61 \pm 15$ & $0.43 \pm 0.09$ & 1,1\\ 

081121 & 2.512 & $-0.46 \pm 0.08\,[-2.19 \pm 0.07]$ & $51.7 \pm 8.3$ & $[8-35\,000]$ & $608 \pm 42$ & $195.4 \pm 33.7$ & 1,1 \\

081203A & 2.10 & $-1.29 \pm 0.14$ & $2.9 \pm 0.2^{\star}$ & $[15-150]$ & $1541 \pm 756$ & $28.3 \pm 8.9$ & 1,1 \\

\textit{081221} & 2.26 & $-0.83 \pm 0.01$ & $24.2 \pm 0.5$ & $[8-35\,000]$ & $284 \pm 2$ & $101 \pm 2$ & 1,3 \\	 

081222 & 2.77 & $-0.90 \pm 0.03\,[-2.33 \pm 0.10]$ & $17.6 \pm 0.58$ & $[8-35\,000]$ & $630 \pm 31$ & $95 \pm 6$ & 1,1 \\ 

090102 & 1.547 & $-0.97 \pm 0.01$ & $29.3 \pm 0.91$ & $[8-35\,000]$ & $1174 \pm 38$ & $45.7 \pm 1.4$ & 1,1 \\ 

\textit{090201} & 2.1 & $-0.97 \pm 0.09\,[-2.80 \pm 0.52]$ & $73.0 \pm 12.6$ & $[20-2000]$ & 489.8 & $269.5 \pm 47.6$ & 1,2 \\

090424 & 0.544 & $-1.02 \pm 0.01\,[-3.26 \pm 0.18]$ & $91.2 \pm 1.4$ & $[8-35\,000]$ & $250.0 \pm 3.4$ & $11.16 \pm 0.18$ & 1,1\\

\textit{090709A} & 1.8 & $-0.85 \pm 0.08\,[-2.7 \pm 0.24]$ & $39 \pm 6$ & $[20-3000]$ & 834.4 & $91.9 \pm 13.9$ & 1,2 \\

090715B & 3.00 & $-1.1 \pm 0.37$ & $9.0 \pm 2.5$ & $[20-2000]$ & $536 \pm 164$ & $82.6 \pm 25.2$ & 1,1 \\

090812 & 2.452 & $-1.03 \pm 0.07$ & $2.77 \pm 0.28^{\star}$ & $[100-1000]$ & $2023 \pm 663$ & $96.3 \pm 16.0$ & 1,1\\
 
090926B & 1.24 & $-0.19 \pm 0.06$ & $4.73 \pm 0.28$ & $[8-35\,000]$ & $212.0 \pm 4.3$ & $4.28 \pm 0.25$ & 1,1\\    

091018 & 0.971 & $-1.53 \pm 0.48$ & $4.32 \pm 0.95$ & $[20-1000]$ & $55 \pm 26$ & $4.75 \pm 1.33$ & 1,1\\
  
091020 & 1.71 & $-1.20 \pm 0.06\,[-2.29 \pm 0.18]$ & $18.8 \pm 2.6$ & $[8-35\,000]$ & $507 \pm 68$ & $32.7 \pm 5.2$ & 1,1 \\   

091127 & 0.49 & $-1.25 \pm 0.05\,[-2.22 \pm 0.01]$ & $93.8 \pm 2.3$ & $[8-35\,000]$ & $51.0 \pm 1.5$ & $9.09 \pm 0.24$ & 1,1\\  

091208B & 1.063 & $-1.29 \pm 0.04$ & $25.6 \pm 0.97$ & $[8-35\,000]$ & $246 \pm 15$ & $17.5 \pm 0.7$ & 1,1\\

\textit{100615A} & 1.4 & $-1.24 \pm 0.07\,[-2.27 \pm 0.11]$ & $8.3 \pm 0.2^{\star}$ & $[8-1000]$ & $206.4 \pm 20.4$ & $10.15 \pm 0.87$ & 1,2 \\    

100621A & 0.542 & $-1.70 \pm 0.13\,[-2.45 \pm 0.15]$ & $17.0 \pm 1.3$ & $[20-2000]$ & $146 \pm 23$ & $3.17 \pm 0.34$ & 1,1 \\

100728B & 2.106 & $-0.90 \pm 0.07$ & $5.43 \pm 0.35$ & $[8-35\,000]$ & $404 \pm 29$ & $18.7 \pm 1.3$ & 1,1 \\  

110205A & 2.22 & $-1.52 \pm 0.14$ & $5.1 \pm 0.7$ & $[20-1200]$ & $715 \pm 238$ & $25.1 \pm 4.3$ & 1,1\\

110503A & 1.613 & $-0.98 \pm 0.08\,[-2.7 \pm 0.3]$ & $100 \pm 10$ & $[20-5000]$ & $572 \pm 50$ & $180.7 \pm 19.7$ & 1,1\\ 


110709A & $-$ & $-1.16 \pm 0.02$ & $15.4 \pm 1.7^{\star}$ & $[10-1000]$ & $533 \pm 37^{\rm b}$ & $-$ & 4,$-$\\    

110709B & $<4$ & $-1.0^{+0.14}_{-0.13}$ & $11 \pm 1$ & $[20-5000]$ & $278^{+43\,\rm b}_{-32}$ & $-$ & 5,6 \\  

110915A & $-$ & $-0.94 \pm 0.23$ & $3.3 \pm 0.2^{\star}$ & $[15-150]$ & $124.8 \pm 41.4^{\rm b}$ & $-$ & 7,$-$\\                 

\bottomrule
\end{tabular}
\end{table*}
\begin{table*}
\scriptsize
\centering
\begin{tabular}{lccccccc}
\toprule
\multicolumn{1}{l}{GRB} &
\multicolumn{1}{c}{$z$} & 
\multicolumn{1}{c}{$\alpha$[$\beta$]} &
\multicolumn{1}{c}{Peak flux} &
\multicolumn{1}{c}{Range} &
\multicolumn{1}{c}{$E_{\rm p}$} &
\multicolumn{1}{c}{$L$} &
\multicolumn{1}{c}{Ref} \\
    &    &                     & $10^{-7}$ erg cm$^{-2}$ s$^{-1}$ &(keV) &(keV) &($\times 10^{ 51}$ erg/s) & \\
    &    &                     &$^\star$ phot cm$^{-2}$ s$^{-1}$ &  &  &  & \\
\midrule

111008A$^{\rm c}$ & 4.989 & $-1.86 \pm 0.09$ & $6.4 \pm 0.7^{\star}$ & $[15-150]$ & 384 & $303.3 \pm 48.8$ & 8,9 \\                                         

111228A$^{\rm c}$ & 0.715 & $-1\,[-2.27 \pm 0.06]$ & $12.4 \pm 0.5^{\star}$ & $[15-150]$ & 46 & $3.64 \pm 0.27$ & 10,11  \\

120102A & $-$ & $-1.19 \pm 0.03$ & $22.8 \pm 1.6^{\star}$ & $[10-1000]$ & $380 \pm 33^{\rm b}$ & $-$ & 12,$-$\\

120116A & $-$ & $-1.31 \pm 0.41$ & $4.1 \pm 0.3^{\star}$ & $[15-150]$ & $-$ & $-$ & 13,$-$\\

120119A & 1.728 & $-0.98 \pm 0.03\,[-2.36 \pm 0.09]$ & $16.9 \pm 0.4^{\star}$ & $[10-1000]$ & $189.2 \pm 8.3$ & $56.9 \pm 2.7$ & 14,15\\                        

120326A & 1.798 & $-1.41 \pm 0.34$ & $4.6 \pm 0.2^{\star}$ & $[15-150]$ & $115 \pm 19$ & $10.8 \pm 1.8$ & 16,17 \\ 

120703A & $-$ & $-0.81^{+0.3}_{-0.25}$ & $54 \pm 13$ & $[20-10\,000]$ & $295^{+88 \, \rm b}_{-56}$ & $-$ & 18,$-$ \\                

120729A$^{\rm c}$ & 0.8 & $-1.62 \pm 0.08$ & $2.9 \pm 0.2^{\star}$ & $[15-150]$ & $192$ & $1.27 \pm 0.11$ & 19,20 \\ 

120802A & 3.796 & $-1.21 \pm 0.47$ & $3.0 \pm 0.2^{\star}$ & $[15-150]$ & $274.3 \pm 93.0$ & $40.7 \pm 5.7$ & 21,22 \\
      
120811C & 2.671 & $-1.4 \pm 0.3$ & $4.1 \pm 0.2^{\star}$ & $[15-150]$ & $157.5 \pm 20.9$ & $25.4 \pm 4.5$ & 23,24 \\  

120907A & 0.97 & $-0.75 \pm 0.25$ & $4.3 \pm 0.4^{\star}$ & $[10-1000]$ & $304.4 \pm 64.8$ & $2.45 \pm 0.27$ & 25,26 \\   

121123A & 2.7 & $-0.96 \pm 0.2$ & $2.6 \pm 0.2$ & $[15-150]$ & $240.5 \pm 18.9$ & $14.9 \pm 1.4$ & 27,28\\

121125A & $-$ & $-1.38 \pm 0.06$ & $4.2 \pm 0.3$ & $[10-1000]$ & $196 \pm 26^{\rm b}$ & $-$ & 29,$-$ \\                               
                  
121209A$^{\rm c}$ & 2.1 & $-1.43 \pm 0.08$ & $3.4 \pm 0.3^{\star}$ & $[15-150]$ & 494 & $17.4 \pm 1.6$ & 30,3\\

130420A & 1.297 & $-1.52 \pm 0.25$ & $3.4 \pm 0.2^{\star}$ & $[15-150]$ & $76.3 \pm 15.6$ & $3.77 \pm 0.65$ & 31,32 \\

130427A & 0.339 & $-0.958 \pm 0.006\,[-4.17 \pm 0.16]$ & $6900 \pm 100$ & $[20-1200]$ & $1371.3 \pm 10.7$ & $384.1 \pm 5.7$ & 33,34\\

130427B$^{\rm c}$ & 2.78 & $-1.64 \pm 0.15$ & $3.0 \pm 0.4$ & $[15-150]$ & 386 & $29 \pm 5$ & 35,36\\                       

130502A & $-$ & $-1.0 \pm 0.3$ & $7 \pm 1$ & $[8-1000]$ & $83 \pm 17^{\rm b}$ & $-$ & 37,$-$\\                                      

130505A & 2.27 & $-0.69 \pm 0.04\,[-2.03 \pm 0.03]$ & $690 \pm 30$ & $[20-1200]$ & $2063.4 \pm 101.4$ & $3959 \pm 172$ & 38,39\\   
      
130527A & $-$ & $1.04 \pm 0.04$ & $500 \pm 30$ & $[20-10\,000]$ & $1380 \pm 120^{\rm b}$ & $-$ & 40,$-$\\       

130606A$^{\rm c}$ & 5.913 & $-1.14 \pm 0.15$ & $2.6 \pm 0.2^{\star}$ & $[15-150]$ & 2032 & $229.5 \pm 24.3$ & 41,42 \\       

130609B & $-$ & $-0.66 \pm 0.22\,[-2.6 \pm 0.2]$ & $13.6 \pm 0.4^{\star}$ & $[10-1000]$ & $491 \pm 20^{\rm b}$ & $-$ & 43,$-$ \\
              
130701A & 1.555 & $-0.9 \pm 0.21$ & $17.1 \pm 0.7$ & $[15-150]$ & $227.9 \pm 31.6$ & $28.9 \pm 1.5$ & 44,45 \\ 

130803A & $-$ & $0.85 \pm 0.09$ & $7.1 \pm 0.3^{\star}$ & $[10-1000]$ & $141.6 \pm 12.2^{\rm b}$ & $-$ & 46,$-$ \\ 

130831A & 0.479 & $-1.61 \pm 0.06\,[-3.3 \pm 0.3]$ & $25 \pm 3$ & $[20-10\,000]$ & $81.3 \pm 5.9$ & $3.68 \pm 0.45$ & 47,48 \\  

130907A & 1.238 & $-0.91 \pm 0.02\,[-2.34 \pm 0.07]$ & $220 \pm 10$ & $[20-10\,000]$ & $881.8 \pm 24.6$ & $185.6 \pm 8.8$ & 49,50 \\

131030A & 1.293 & $-0.71 \pm 0.12\,[-2.95 \pm 0.28]$ & $100 \pm 10$ & $[20-10\,000]$ & $405.9 \pm 22.9$ & $103 \pm 11$ & 51,52\\

131105A & 1.686 & $-0.88 \pm 0.38\,[-2.33 \pm 0.33]$ & $20 \pm 2$ & $[20-10\,000]$ & $419 \pm 102$ & $37.4 \pm 4.9$ & 53,54\\  

140102A & $-$ & $-0.71 \pm 0.02\,[-2.49 \pm 0.07]$ & $49.7 \pm 0.5^{\star}$ & $[10-1000]$ & $186 \pm 5^{\rm b}$ & $-$ & 55,$-$ \\  

140206A & 2.73 & $-1.04 \pm 0.15$ & $19.4 \pm 0.5^{\star}$ & $[15-150]$ & $376.4 \pm 54.1$ & $141.5 \pm 4.8$ & 56,57 \\

140215A & $-$ & $-0.66 \pm 0.11\,[-2.94 \pm 0.35]$ & $35.7 \pm 3.5$ & $[20-10\,000]$ & $214 \pm 14^{\rm b}$ & $-$ & 58,$-$ \\    

140419A & 3.956 & $-0.63^{+0.36}_{-0.22}\,[-2.3^{+0.4}_{-2.5}]$ & $47^{+18}_{-19}$ & $[20-10\,000]$ & $1452.1 \pm 416.3$ & $572.6 \pm 25.2$ & 59,60 \\

140506A & 0.889 & $-0.9 \pm 0.2\,[-2.0 \pm 0.1]$ & $14.2 \pm 0.7^{\star}$ & $[10-1000]$ & $266 \pm 68$ & $11.5 \pm 1.3$ & 61,62\\

140512A & 0.72 & $-1.33 \pm 0.03$ & $11.0 \pm 0.3^{\star}$ & $[10-1000]$ & $1011 \pm 145$ & $5.28 \pm 0.47$ & 63,64 \\ 

140619A & $-$ & $-1.45 \pm 0.14$ & $4.6 \pm 0.2^{\star}$ & $[15-150]$ & $117.8 \pm 46.2^{\rm b}$ & $-$ & 65,$-$ \\ 

140628A & $-$ & $-1.56 \pm 0.09$ & $2.8 \pm 0.2^{\star}$ & $[15-150]$ & $-$ & $-$ & 66,$-$\\

140629A & 2.275 & $-1.42 \pm 0.54$ & $4.7 \pm 0.7$ & $[20-10\,000]$ & $281.7 \pm 57.4$ & $27.1 \pm 5.5$ & 67,68\\

140703A & 3.14 & $-1.10 \pm 0.06$ & $4.1 \pm 0.2^{\star}$ & $[10-1000]$ & $732.8 \pm 58.0$ & $41.6 \pm 2.2$ & 69,70\\

\bottomrule
\end{tabular}
\end{table*}

Then, we extended the BAT6 revised sample with the new events detected since 2012 which satisfy the observability criteria of Jackobsson et al. (2006). The extended sample contains 99 GRBs. We collected the spectral parameters of the new GRBs in the existing literature. For those events (6/41 bursts) with only a \sw\ single power law spectrum,  we estimated \ep\ through the Sakamoto et al. (2009) relation. The BAT6 extended (BAT6ext, hereafter) counts 82/99 GRBs with $z$ ( and  81/99 with $z$ and $L$. Its completeness in redshift is $\sim 82\%$.

The BAT6ext is presented in Tab. \ref{tab1}. The first 58 bursts are the original BAT6, while the others constitute the extension. For each GRB, Tab. \ref{tab1} shows the redshift $z$, the spectral parameters (high and low photon indices $\alpha$ and $\beta$ and the rest frame peak energy $E_{\rm p}^{\rm rest}$), the peak flux with the relative energy band, the isotropic equivalent  luminosity $L$. The spectrum is a cut-off power-law (CPL) if only the low energy photon index $\alpha$ is reported and a Band function if also the high energy photon index $\beta$ is given. When $z$ is not measured, we report the observed peak energy. The luminosities reported in the table are calculated in the $[1-10^4]$ keV rest frame energy range only for those GRBs having both $z$ and $E_{\rm p}$.

\section{Luminosity function and GRB formation rate}
\label{sec5}

In this section we will apply the $C^{-}$ method as originally proposed by Lynden-Bell et al. (1971) and applied to GRBs by e.g. Yonetoku et al. (2004,2014), Kocevski \& Liang (2006), Wu et al. (2014). This method is based on the assumption that the luminosity is independent from the redshift. However, as discussed in Petrosian et al. (2015) a strong luminosity evolution could be present in the GRB population. Efron \& Petrosian (1992) proposed a non--parametric test to estimate the degree of correlation of the luminosity with redshift induced by the flux in a flux--limited sample. This is also the case of the BAT6ext sample and the first step will be to quantify the degree of correlation. Yu et al. (2015) and Petrosian et al. (2015) indeed found that the luminosity evolves with redshift within their samples as $(1+z)^{2.43^{+0.41}_{-0.38}}$ (Y15) or $(1+z)^{2.3\pm0.8}$ (P15). We applied the same method  of Y15 and P15 (also used in Yonetoku et al. 2004,2014) to the BAT6ext sample: we define the luminosity evolution $L=L_{0}(1+z)^k$ (as in Y15), where $L_{0}$ is the de-evolved luminosity, and compute the modified Kendall correlation coefficient (as defined in Efron \& Petrosian 1992). We find, consistently with the results of Y15 and P15, $k=2.5$. Similar results were obtained through the indirect method (see \S2) by S12 using the BAT6 sample. 

We can now define the de-evolved luminosities $L_{0} = L/(1+z)^k$ for every GRBs and apply the Lynden--Bell $C^{-}$ method to derive the cumulative luminosity function $\Phi(L_{0})$ and the GRB formation rate \gfr.

\begin{figure*}
\centering
\hskip -0.7truecm
\includegraphics[width=9.truecm]{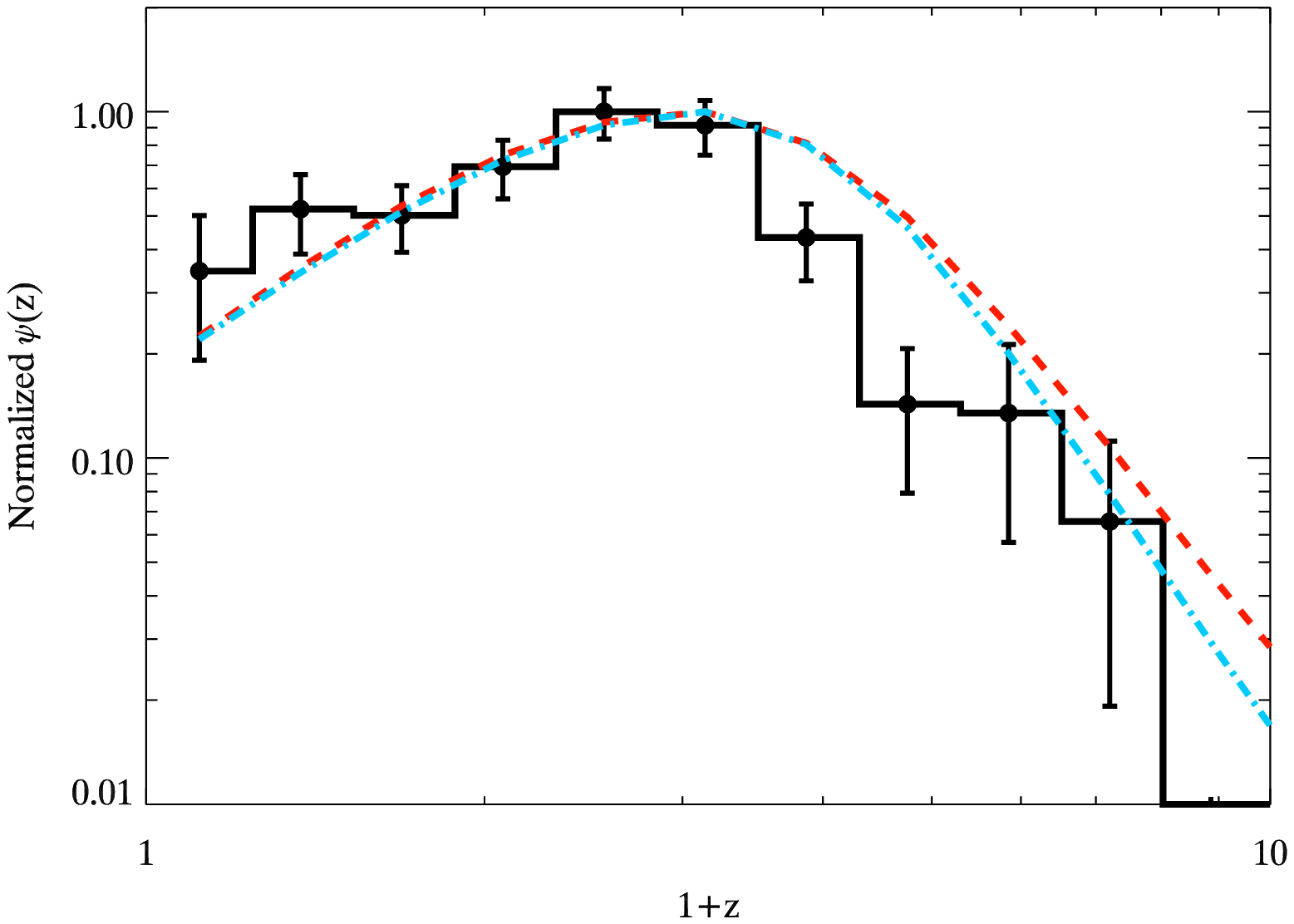}
\includegraphics[width=9.truecm]{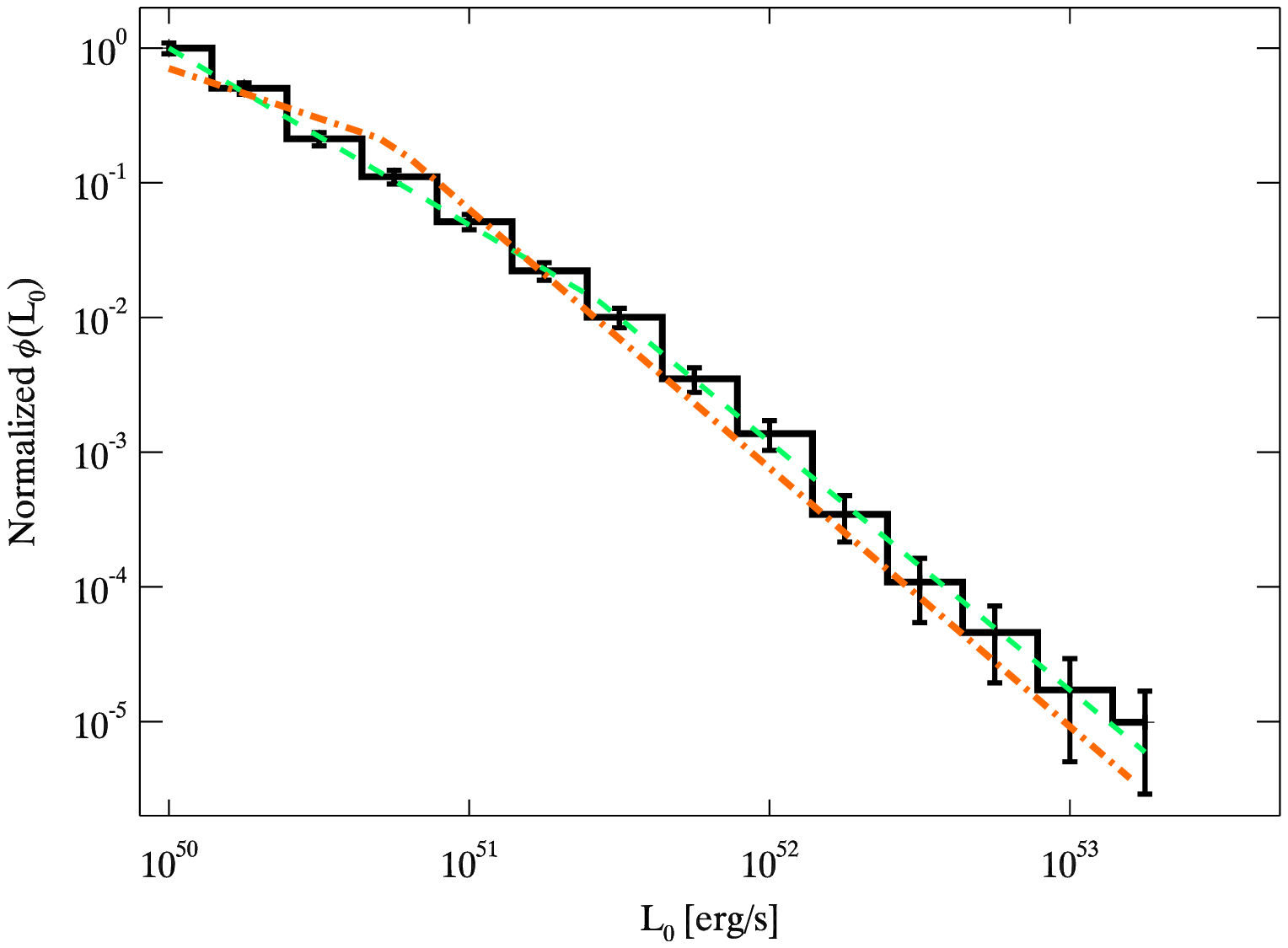}
\caption{\label{PhiR}{\it Left panel:} GRB formation rate $\psi(z)$ obtained with the $C^{-}$ method using the BAT6ext sample (black solid line). The dashed red line and the dot-dashed cyan line are the SFR models of Hopkins \& Beacom (2006) and Cole et al. (2001) shown here for reference. All the curves are normalised to their maxima. {\it Right panel}: luminosity function $\phi(L_{0})$ obtained with the $C^{-}$ method using the BAT6ext sample (black solid line). The best fit model describing this function is a broken power-law (dashed green line) with ($a=-1.32 \pm 0.21$, $b=-1.84 \pm 0.24$, $L_{\rm b}=10^{51.45 \pm 0.15}$ erg/s). The orange dot-dashed line is the luminosity function obtained by S12 in the case of pure luminosity evolution.}
\end{figure*} 

For the ith GRB in the BAT6ext sample, described by its ($L_{\rm 0,i}, z_{\rm i}$), we consider the subsample $J_{\rm i}=\left\lbrace j | L_{\rm 0,j} > L_{\rm 0,i} \wedge z_{\rm j} < z_{\rm max,i} \right\rbrace$ and call $N_{\rm i}$ the number of GRBs it contains.
Similarly, we define the subsample $J_{\rm i}' = \left\lbrace j | L_{\rm 0,j} > L_{\rm lim,i} \wedge z_{\rm j} < z_{\rm i} \right\rbrace$ and we call $M_{\rm i}$ the number of GRBs it contains. $L_{\rm lim,i}$ is the minimum luminosity corresponding to the flux limit $S$ of the sample at the redshift $z_{\rm i}$. $z_{\rm max,i}$ is the maximum redshift at which the {\it i-}th GRBs with luminosity $L_{\rm 0,i}$ can be observed (i.e. with flux above the limit $S$).  $L_{\rm lim}$ and $z_{\rm max}$ are computed applying the K-correction because the limiting flux is computed in the observer frame \sw/BAT $[15-150]$ keV energy band.

Through $M_{\rm i}$ and $N_{\rm i}$ we can compute the cumulative luminosity function $\Phi(L_{0})$ and the cumulative GRB redshift distribution $\zeta(z)$: 

\begin{equation}
\Phi(L_{\rm 0,i}) = \prod_{\rm j<i} (1+\frac{1}{N_{\rm j}})
\label{PhiLF}
\end{equation}

and

\begin{equation}
\zeta (z_{\rm i}) = \prod_{\rm j<i} (1+\frac{1}{M_{\rm j}})
\label{psi}
\end{equation}

From the latter we can derive the GRB formation rate as:

\begin{equation}
\psi(z)=\frac{d\zeta(z)}{dz}(1+z)\left[\frac{dV(z)}{dz}\right]^{-1}
\label{rate}
\end{equation}
  
where $dV(z)/dz$ is the differential comoving volume. The differential luminosity function $\phi(L_{0})$ is obtained deriving the cumulative one $\Phi(L_{0})$.

The functions $\phi(L_{\rm 0})$ and $\psi(z)$ are shown in Fig. \ref{PhiR}. Errors on \flumdev\ are computed propagating the errors on the cumulative one assuming Poisson statistics.  The errors on the \gfr\ are computed from the number $n$ of GRBs within the redshift bin. We assume that the relative error $\epsilon=1/\sqrt{n}$ is the same affecting \gfr.

\section{Results}

The luminosity function $\phi(L_{\rm 0})$ obtained with the BAT6ext sample is shown in Fig. \ref{PhiR} (right panel) by the black symbols. Data are normalised to the maximum. The best fit model (green dashed line in Fig. \ref{PhiR} - right panel) is  represented by a broken power law function with $a=-1.32 \pm 0.21$, $b=-1.84 \pm 0.24$, $L_{\rm b}=10^{51.45 \pm 0.15}$ erg/s (where $a$ and $b$ represent the slopes of the power law below and above $L_{\rm b}$ - $\chi^{2}/{\rm d.o.f.} = 0.47$). In Fig. \ref{PhiR} we also show, for comparison, the luminosity function derived in S12 through the indirect method described in \S2 in the case of pure luminosity evolution ($a = -0.74 _{-1.42} ^{+1.36}$, $b = -1.92 _{-0.14} ^{+0.11}$, $L_{\rm b} = 5.5 ^{+6.9} _{-3.4} \times 10^{50}$ erg/s). This model is consistent with the result obtained in our analysis. 
 
The GRB formation rate \gfr\ obtained with the BAT6ext is shown by the black symbols in Fig. \ref{PhiR} (left panel). The green dashed and the cyan dot-dashed lines are, respectively, the SFR of Hopkins \& Beacom (2006) and Cole (2001). Data and models are normalised to their peak. Contrary to the results reported in P15 and Y15, the \gfr\ that we derive increases up to $z\sim 2$. This trend is consistent with the independent estimates obtained through host galaxy studies.  

The BAT6ext sample has a smaller completeness in redshift ($\sim 82\%$) with respect to the revised BAT6 ($\sim 97\%$). We checked if this could in some way modify the shape of $\psi(z)$. For this reason we also computed the \gfr\ using only the 56 objects of the revised BAT6 sample which turns out to be slightly steeper both at low and high redshifts than the one obtained with the BAT6ext but totally consistent within the errors. We conclude that the lower completeness in redshift of the BAT6ext does not introduce any strong bias in the \gfr\ and \flumdev\ obtained.

\begin{figure*}
\centering
\hskip -1truecm
\includegraphics[width=9.2truecm]{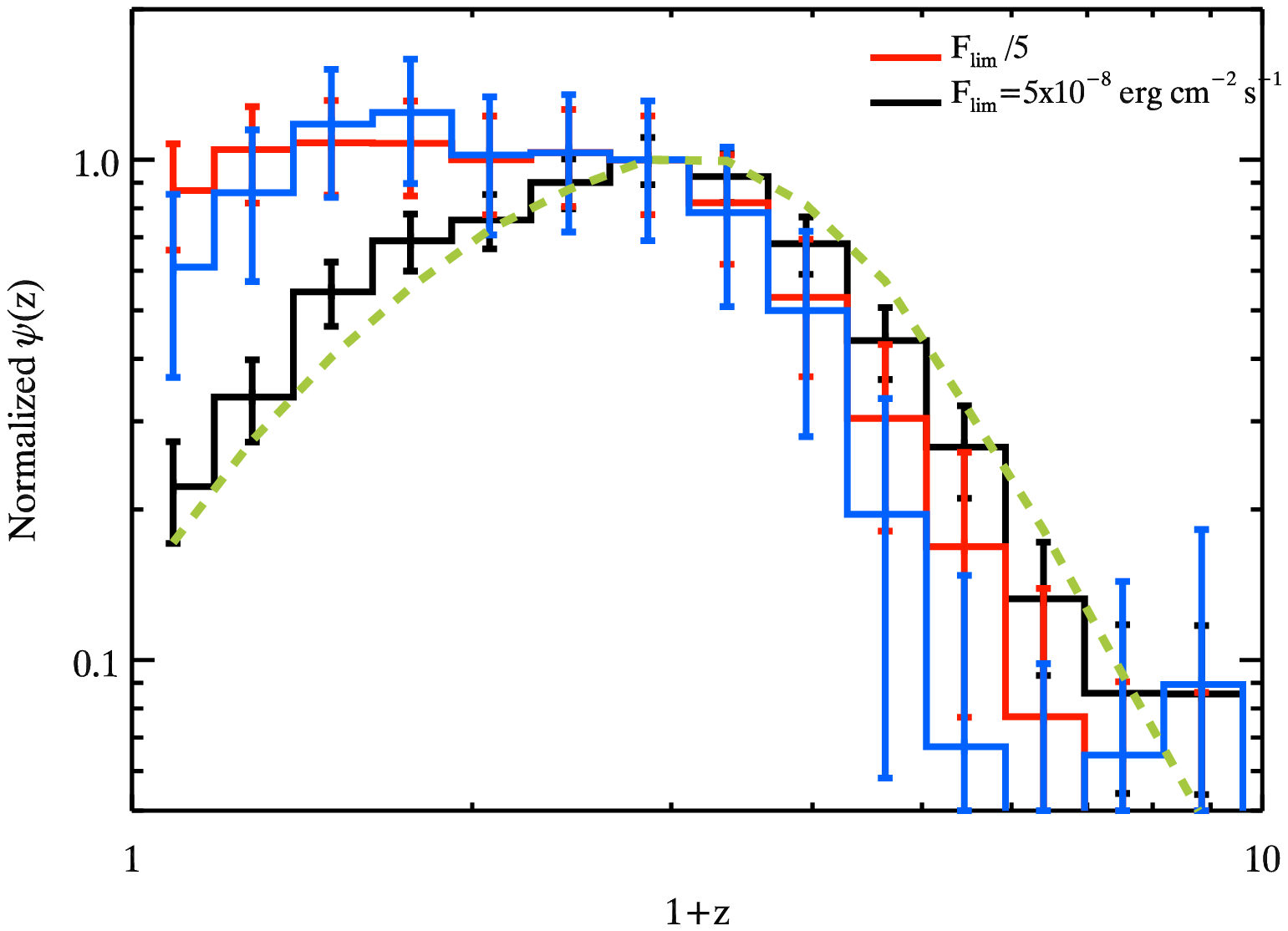} 
\includegraphics[width=9.2truecm]{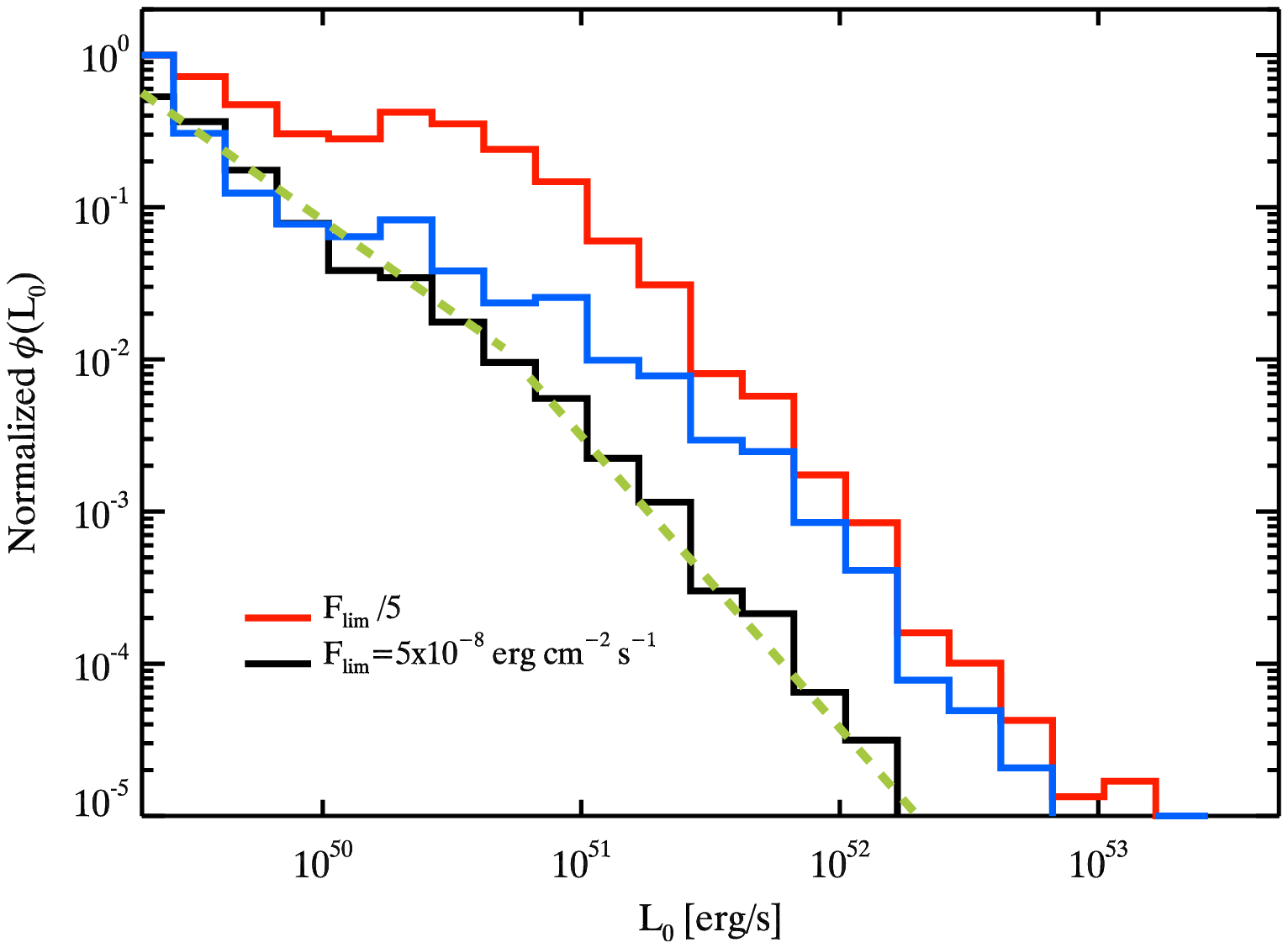} 
\caption{\label{fig4}{\it Left:} GRB formation rate (normalised to its peak) for the simulated population of GRBs with flux limit $5\times10^{-8}$ erg cm$^{-2}$ s$^{-1}$ (black symbols). The GRB formation rate assumed in the simulation is  shown by the dashed green line. The red symbols show the results obtained from the same sample using, for the analysis, a flux limit a factor of 5 smaller than the real one. Blue symbols are obtained by mimicking the sample incompleteness by removing randomly some GRBs near the flux threshold adopted for the sample selection. {Right panel:} cumulative luminosity function, normalised to the first bin. The black, red and blue symbols are the same as for the left panel. The assumed luminosity function is shown by the dashed green line.}
\end{figure*}     

\section{Monte Carlo test of the $C^{-}$ method}
\label{sec6}
We now test the $C^{-}$ method used to derive \gfr\ and \flumdev. Through a Monte Carlo simulation (similar to e.g. Ghirlanda et al. 2015) we explore how well the method adopted in \S\ref{sec5} can recover the input assumptions, i.e. $\psi(z)$ and LF $\phi(L)$. In particular, we will show that if the sample used is highly incomplete the resulting GRBFR and LF can differ significantly from the input ones. In particular, incomplete samples (either in flux and/or redshift) may produce a misleading excess of low redshift GRBs with respect to the assumed \gfr. 

We simulate GRBs distributed in redshift according to the GRB formation rate $\psi(z)$ of Li 2008 (see also Hopkins \& Beacom 2008):
\begin{equation}
\psi(z)=\frac{0.0157+0.118z}{1+(z/3.23)^{4.66}}
\label{z}
\end{equation}
where $\psi(z)$, in units of ${\rm M}_{\odot}$ ${\rm yr}^{-1}$ ${\rm Mpc}^{-3}$, represents the formation rate of GRBs and we assume it can extend to $z \le 10$. We stress that for the scope of the present test any other functional form of $\psi(z)$ could be assumed. 

We adopt a luminosity function $\phi(L)$ as obtained by Salvaterra et al. 2012 from a complete sample of Swift GRBs:
\begin{equation}
\phi(L) \propto
\begin{cases}
\left(L\over L_{\rm b}\right)^{a}  & \text{, $L \leq L_{\rm b}$} \\
\left(L\over L_{\rm b}\right)^{b}   & \text{, $L> L_{\rm b}$}
\end{cases}
\label{lf}
\end{equation}
composed by two power-laws with a break at $L_{\rm b}$. We adopted arbitrary parameter values: $a=-1.2$, $b=-1.92$ and $L_{\rm b}=5\times10^{50}$ erg s$^{-1}$. We further assume an evolution of the luminosity proportional to $(1+z)^{k}$  with $k=2.2$. We assume this is the bolometric luminosity of the simulated bursts and compute the corresponding bolometric flux. Also for \flum\ we use this functional form but any other function could be assumed for the scope of the present test.

With these two assumptions we simulate a sample of GRBs with a flux limit $F_{\rm lim}=5\times10^{-8}$ erg cm$^{-2}$ s$^{-1}$ and we analyse it with the $C^{-}$ method. Accounting for the truncation of the sample we recover through the statistical method  of Efron \& Petrosian (1992) the luminosity evolution in the form $(1+z)^{k}$, with $k \sim 2.2$. 

Then we work with the de-evolved GRB luminosities $L_{0}=L/(1+z)^{2.2}$ and derive the GRB formation rate $\psi(z)$ and the luminosity function $\phi(L_{0})$ through the $C^{-}$ method proposed by Lynden-Bell et al. (1971). The left panel of Fig. \ref{fig4} shows (black symbols) that we recover the GRB formation rate of Eq. \ref{z} (dashed green line) that we adopted in the simulation. Similarly the right panel of Fig. \ref{fig4} shows that we also recover the  luminosity function  that we adopted in the simulated sample (Eq. \ref{lf} - shown by the dashed green line in Fig. \ref{fig4}). 

We then tested what happens if we apply the same method to an incomplete sample. Firstly we applied the $C^{-}$ method to the same simulated sample (which is built to be complete to $F_{\rm lim} = 5\times10^{-8}$ erg cm$^{-2}$ s$^{-1}$) from which we removed randomly a fraction of the bursts close to $F_{\rm lim}$. This new sample is clearly incomplete to $F_{\rm lim}$. The results are shown in Fig. \ref{fig4} by the blue symbols. We find that the GRB formation rate $\psi(z)$ is flat at low redshifts (i.e. below $z=2$) showing a clear excess with respect to the assumed function (cf blue symbols with the green dashed line in the left panel of Fig. \ref{fig4}). The luminosity function is flatter than the assumed one (cf blue symbols with the green dashed line in Fig. \ref{fig4} right panel).

Similar results are obtained by assuming for the derivation of \gfr\ and \flumdev\ a flux limit which is a factor of 5 smaller than that used to construct the simulated sample. This is another way to make the sample artificially incomplete. The results are shown by the red symbols in the panels of Fig. \ref{fig4}. Note that in this second test the sample used is the same but it is analysed through the $C^{-}$ method assuming it is complete with respect to a flux limit which is smaller (a factor of 5) than the one corresponding to its real completeness (i.e. $5\times10^{-8}$ erg cm$^{-2}$ s$^{-1}$).

These simulations show that if the samples adopted are highly incomplete in flux, one obtains an excess at the low redshift end of the GRB formation rate and a flatter luminosity function. 

\section{Summary and discussion}

The aim of this work is to derive the luminosity function of long GRBs and their formation rate. To this aim we apply a direct method (Lynden--Bell et al. 1971) and its specific version already applied to GRBs, e.g. Yonetoku et al. (2004, 2014), Kocevski \& Liang (2006), Wu et al. (2014), P15, Y15. This is the first time this method is applied to a well defined sample of GRBs which is complete in flux and 82\% complete in redshift. 

We build our sample of long GRBs starting from  the BAT6 complete sample (Salvaterra et al. 2012): this was composed by 58 GRBs detected by the \sw\ satellite satisfying the multiple observational selection criteria of Jackobsson (2006) and having a peak photon flux  $P \ge 2.6$ ph cm$^{-2}$ s$^{-1}$. Here, we update the redshift measurement of 8 GRBs of the BAT6 (marked in italics in Tab. \ref{tab1}) and accordingly revise their luminosities. Then, we update this sample to GRB\,140703A ending with 99 objects. We collect their redshift measurements and spectral parameters from the literature (see Tab. \ref{tab1}). The BAT6ext sample has a redshift completeness of $\sim 82\%$ (82/99 burst with $z$ measured) and counts 81/99 bursts with well determined $L$. 

We analyzed the BAT6ext sample searching for a possible luminosity evolution induced by the flux threshold through the method proposed by Efron \& Petrosian (1992). We find that the $L-z$ correlation introduced by the truncation due to the flux limit can be described as $L = L_{0}(1+z)^k$ with $k=2.5$. This result is in agreement with what found by other authors (Yonetoku et al. 2004, Wu et al. 2012, P15, Y15).
With the BAT6ext sample after de-evolving the luminosities for their redshift dependence we find that:

\begin{itemize}
\item the luminosity function $\phi(L_{0})$ is a monotonic decreasing function well described by a broken powerlaw with slopes $a=-1.32 \pm 0.21$ and $b=-1.84 \pm 0.24$ below and above, respectively, a characteristic break luminosity $L_{\rm b}=10^{51.45 \pm 0.15}$ erg/s (green dashed line in the right panel of Fig. \ref{PhiR}). This result (shape, slopes and characteristic break) is consistent with the luminosity function found by S12 (orange dot-dashed line in Fig. \ref{PhiR} - right panel). 

\item The cosmological GRB formation rate $\psi(z)$ (black solid line in the left panel of Fig. \ref{PhiR}) increases from low redshifts to higher values peaking at $z \sim 2$ and decreases at higher redshifts. This trend is consistent with the shape of the SFR of Hopkins \& Beacom (2006) and Cole et al. (2001) (color lines in Fig. \ref{PhiR}). Our results on $\psi(z)$ is in contrast with the GRBFR recently found by P15 and Y15, who, applying the same method to differently selected GRB samples, report the existence of an excess of low redshift GRBs. 

\end{itemize}

However, the result that GRBs evolve in luminosity is not the probe that GRBs had experienced a pure luminosity evolution. In fact the $C^{-}$ method assumes the independence between $L$ and $z$ and the non parametric method of Efron \& Petrosian used to get the de-evolved luminosities assigns all the evolution in the luminosity. For this reason we are not able to distinguish between an evolution in luminosity or in density (see also Salvaterra et al. 2012). The possible density evolution case requires the investigation of the applicability of similar methods to the GRB samples and will be the subject of a forthcoming work (Pescalli et al. , in preparation). 

Finally, intrigued by the different results with respect to Y15 and P15, we performed Monte Carlo simulations in order to test the robustness of the $C^{-}$ method. We showed that this method can correctly recover the LF \flumdev\ and the GRBFR $\psi(z)$ assumed in the simulation only if the sample of GRBs it is applied to is complete in flux and has a high level of completeness in redshift. Using incomplete samples or a sample incomplete in redshift, the resulting GRBFR and LF can be different from the assumed ones. Indeed, this could account for the excess of the rate of GRBs at low redshift as recently reported. 

\begin{acknowledgements}
We acknowledge the 2011 PRIN-INAF grant for financial support. We acknowledge the financial support of the UnivEarthS Labex program at Sorbonne Paris Cit\'e (ANR-10-LABX-0023 and ANR-11-IDEX-0005-02). GG thanks the Observatoire de Paris (Meudon) for support and hospitality during the completion of this work. We thank S. Campana and G. Tagliaferri for useful discussion. 
\end{acknowledgements}

\appendix
\section{The redshift integrated luminosity function}
\label{sez4}

In this section we show how we derive the luminosity function of GRBs {\it integrated over all the redshift space} using the BAT6ext sample. This is not the typical luminosity function which is derived through indirect methods, it is free from any functional form and only uses the $1/V_{\rm max}$ concept, i.e. the maximum comoving volume within which a real burst with an observed luminosity could be detected by a given instrument. This is a generalisation of the $<V/V_{\rm max}>$ method proposed by Schmidt (1968) and applied to quasars (see Avni \& Bachall 1980 for an exhaustive description).

However, the luminosity function obtained with this method is not the canonical \flum\ but it is the result of the integration of the latter convolved with the GRB formation rate $\psi(z)$ over $z$. We call this function the convolved luminosity function $\Sigma(L)$ (CLF). In principle the luminosity function could evolve with redshift ($\phi(L,z)$). For this reason the shape of the CLF could be different from the shape of the canonical $\phi(L)$. It would be simply proportional to $\phi(L)$ only if the luminosity function does not evolve, either in luminosity or in density. 
Indeed, it is possible to express $\Sigma(L)$ in terms of luminosity function and GRB formation rate:

\begin{equation}
\Sigma(L) = \int _{0}^{\infty} \phi(L,z)\psi(z) \, dz
\label{Xi}
\end{equation}

The advantage is that it can be obtained directly from the data and it is extremely robust if derived through a flux limited sample. Any evolution with redshift of the luminosity or density is in this CFL. We use the 81/99 GRB with both $z$ and determined $L$ reported in Tab.1.  

\begin{figure}
\centering
\includegraphics[width=9truecm]{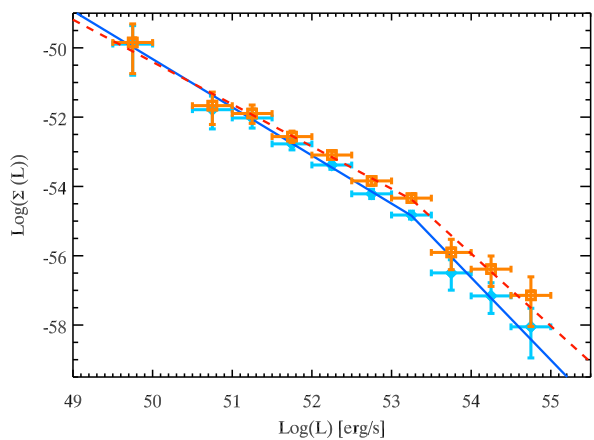}
\caption{\label{trueCLF}{The blue filled points represent the observed CLF $\Sigma(L)$ while the orange squares represent the true CLF $\tilde{\Sigma}(L)$. The solid blue line ($a = -1.39 \pm 0.14$, $b = -2.38 \pm 0.41$, $L_{\rm b} = 10^{53.3 \pm 0.3}$ erg/s) and the dashed orange line ($a = -1.22 \pm 0.1$, $b= -2.09 \pm 0.95$, $L_{\rm b} = 10^{53.3 \pm 1}$ erg/s) are the best fit models of the observed and true CLF respectively.}}
\end{figure} 

 
For each GRB of the BAT6ext sample with an associated $L$ we estimate the maximum volume $V_{\rm max}$ within which the burst could still be detected because its flux would be larger than the chosen threshold i.e. $P_{\rm lim} = 2.6$ ph cm$^{-2}$ s$^{-1}$ in the 15--150 keV energy band. The observed photon flux in the \sw/BAT energy band as a function of the varying redshift is: 
\begin{equation}
P(z) = \frac{L}{4\pi d_{\rm L}^2(z)}\frac{\int ^{150 keV}_{15 keV} N(E) \, dE}{{\int ^{10^4/(1+z) keV} _{1/(1+z) keV} EN(E) \, dE}}  
\label{phflux}
\end{equation}
where $N(E)$ is the observed photon spectrum of each GRB and $d_{\rm L}(z)$ is the luminosity distance at redshift $z$. The extremes of the integral in the denominator correspond to the same values adopted to compute $L$. The maximum redshift $z_{\rm max}$ corresponds to the redshift satisfying $P(z_{\rm max}) = P_{\rm lim}$. 

Considering the typical \sw/BAT field of view $\Omega = 1.33$ sterad and the time of activity of \sw\ $T \sim 9$ yr that covers the BAT6ext sample, we can compute the rate $\rho_{\rm i} = 4\pi/\Omega TV_{\rm max,i}$ for each GRB. We divided the observed range of luminosities in bins of equal logarithmic width $\Delta$ and estimate $\Sigma_{\rm j}(L_{\rm j})$ for each bin as 
\begin{equation}
\Sigma_{\rm j}(L_{\rm j}) = \frac{1}{\Delta L_{\rm j}} \sum \rho _{\rm i}
\label{Xi}
\end{equation} 
where the sum is made over the bursts with luminosities $L{\rm j}-\Delta/2\le L_{\rm i} \le L{\rm j}+\Delta/2$. The discrete convolved luminosity function is shown in Fig. \ref{trueCLF}. The normalisation is obtained also considering that the bursts in the BAT6ext sample represent approximately 1/3 of the total number of \sw\ detected GRBs with peak flux P$\ge$2.6 ph cm$^{-2}$ s$^{-1}$. We verified that the BAT6ext sample (99 objects) is representative in terms of peak flux distribution of the larger population. The error bar associated to the discrete CLF are mainly related to the Poissonian error on the objects counting within the luminosity bin (see also e.g. Wolter et al. 1994). 
 

When computing individual rates, we have used the mission elapsed time $T$. However, this is the observer frame time and the rate should be corrected for the cosmological time dilation. This means that, at higher $z$, the same subset of sources should occur with a larger frequency. Therefore, we average out the elapsed time on the redshift interval $[0,z_{\rm max}]$
\begin{equation}
<T> = \frac{\int ^{z_{\rm max}} _{0} \frac{T}{1+z} \, dz}{\int ^{z_{\rm max}} _{0} \, dz}
\label{meant}
\end{equation}
and use it in the computation of Eq. \ref{Xi}. As shown in Fig. \ref{trueCLF}, the true CLF $\tilde{\Sigma}(L)$ is flatter than the observed one because, in general, the true elapsed time is smaller than the observed one. Moreover, the correction is more pronounced for high luminosity GRBs observable up to high redshifts. 

The CLF obtained with the extended BAT6ext sample is shown in Fig. \ref{trueCLF} by the filled symbols. The observed CFL (blue symbols in Fig. \ref{trueCLF}) can be adequately represented by a broken power law function (solid blue line in Fig. \ref{trueCLF}) with slopes $-1.39 \pm 0.14$ and $-2.38 \pm 0.41$ below and above the break luminosity $L_{\rm b} = 10^{53.3 \pm 0.3}$ erg/s ($\chi^{2}/{\rm d.o.f.} = 1.04$). When we correct for the cosmological time dilation the true CLF appears slightly flatter (slopes $-1.22 \pm 0.1$ and $-2.09 \pm 0.95$ below and above the break $L_{\rm b} = 10^{53.3 \pm 1}$ erg/s  - $\chi^{2}/{\rm d.o.f.}= 0.99$). 

We can look at this function as a redshift integrated luminosity distribution. This is the most direct information that we can obtain from data, in fact, it is obtained without any type of assumption or observational constraints. Similarly to the flux distribution $\log N- \log S$ and to the redshift distribution it can be used as a constrain. The LF and the GRBFR obtained with other methods should reproduce this CLF once convolved together and integrated over $z$.

\end{document}